\documentclass[11pt,twoside]{article}
\usepackage{asp2010}
\usepackage{natbib}

\aspvolume{471} 
\aspvoltitle{Origins of the Expanding Universe: 1912-1932}
\aspcpryear{2013} 
\aspvolauthor{Michael J. Way and Deidre Hunter, eds.} 

\resetcounters

\begin{document}

\title{Henry Norris Russell and the Expanding Universe}
\author{David DeVorkin\affil{Senior Curator, National Air and Space Museum, Smithsonian
Institution, Washington D.C., USA}}

\begin{abstract}
Henry Norris Russell, one of the most influential American astronomers of the
first half of the 20th Century, had a special place in his heart for the Lowell
Observatory. Although privately critical of the founder for his pronouncements
about life on Mars and the superiority of the Mars Hill observing site, he
always supported the Observatory in public and professional circles.
He staunchly supported Tombaugh's detection of a planet as leading from
Lowell's prediction, and always promoted V.~M. Slipher's spectroscopic
investigations of planetary and stellar phenomena. But how did he react to
Slipher's puzzling detection of the extreme radial velocities of spiral nebulae
starting in 1912, and how did he regard the extension and interpretation of
those observations by Hubble and others in following decades? Here we describe
the arc of Russell's reactions, dating from Slipher's first detection, as an
indicator of how mainstream stellar astronomers reacted to the concept of an
expanding universe.
\end{abstract}

\section{Introduction}
Henry Norris Russell was one of the most influential American astronomers of the first half of the 20th
Century \citep{2000hnr..book.....D}. During his most active years, dating
roughly from the eve of the first World War and lasting till the eve of the
second,  his influence was deeply felt not only as a frequent referee and
arbiter of standards for American professional journals relating to astronomy,
but as a spokesman for the profession, reaching popular and technically informed
audiences in a monthly astronomy column for \emph{Scientific American}.  His column
provided news of upcoming celestial events  surrounding a simple star map, but
it also became a bully pulpit for promoting the deeds and discoveries of
astronomers worldwide, as did his two-volume textbook that was an introduction
to the profession for over a generation of astronomers \citep{1938Rus..book.....R}.

Russell's career spanned what Peter van de Kamp, Richard Berendzen and others
have called, first, the ``Galactocentric Revolution"  and then the
``Acentric Revolution" \citep{1965PASP...77..325V, 1975VA.....17...65B}.
History will remember the 20th Century as the period when these revolutions in
the human comprehension of the universe, in conjunction with the establishment
of relativistic cosmology by Einstein and others, were discovered, elaborated and
reinforced.\footnote{An excellent summation of present historical knowledge,
with an extensive bibliography, can be found in \cite{2009JHA....40...71S}.}

Today, these concepts are in place, but how did those who were born and raised
in a static stellar universe react at the time?  Observational astronomers rose
to the challenge, pragmatically conducting observational tests suggested by
Einstein or inferred from his theories.\footnote{There is an extensive literature, from,
for instance, \cite{1966eubg.book.....W} to \cite{1966eubg.book.....C}.}
Those without the means to make the delicate observations,
or without the inclination, however,
had fewer options to participate: among them, either ignore the implications and
continue on studying the stellar universe, or incorporate the implications when
they might be useful. 

Russell was in the last category.  The Astrophysics Data System suggests that
Russell invoked the term ``expanding universe" only twice in refereed publications
from 1931 - 1940, less frequently than de Sitter, C.~A. Chant, W.~H. McCrea and
G.~C. McVittie (9 times each), Eddington (4 times) and Hubble
(3 times).\footnote{There were 224 hits total using the Astrophysics Data System's
full-text search (Beta version), searching on refereed papers with synonyms
disabled.}
And distinct from the others (save Chant, who was acting in his capacity as
editor of the \emph{Journal of The Royal Astronomical Society of Canada}),
Russell did not try to contribute to the new dynamics other than to
speculate on its implications for the formation of planetary systems. 

For these reasons, recounting how Russell reacted first to V.~M. Slipher's
observations of the extreme velocities of the spiral nebulae and then to their
elaboration and correlation with distance by Hubble and others, will offer
insight into how ``normal" astronomers engaged the new framework. Of course,
following Russell alone will not provide a full historical picture, but
considering his influence, his example is a good place to start, on the occasion
of this celebration of the centennial of Slipher's initial detection of what
ultimately was interpreted as the expansion of the Universe.

\section{First Reactions}

In April 1920, the Dearborn Observatory astronomer Philip Fox wrote his friend V.~M.
Slipher, congratulating him on receiving the 1919 Lalande Prize of the Paris Academy for his
nebular work.  Fox also took the occasion to recall for Slipher how astronomers reacted when, at
the 16th meeting of the American Astronomical Society (AAS) in Atlanta
in December 1913, he read one of Slipher's first
announcements of the great radial velocities of a few bright spiral nebulae.  ``It has been a
matter of very great gratification to me to see the recognition which your splendid work has been
receiving," \cite{Fox-20} began, but then pointedly recalled:

\begin{quote}
I believe you remember that at the time of the Atlanta meeting of the
Astronomical Society I presented your paper on the high velocity of the nebulae,
and it was greeted with some expression of incredulity, especially on the part of
Professor Russell, to which I made a heated rejoinder...  I note that he makes a
confident reference to your work, however, in ``Some Problems of Sidereal
Astronomy," published by the National Research Council.
\end{quote}

In the some four years between Russell's initial reaction to Slipher's work and when he prepared
his influential essay ``Some Problems of Sidereal Astronomy," he came to accept the velocities as
measured quantities.   As his essay was intended to be a referendum on the state of astronomy,
pointing to work that had to be done, he clearly identified the velocities of the spirals as a
problem that needed to be addressed in the context of the ``Sidereal Problem."  He, along with J.
C. Kapteyn, A. Eddington and most mainstream astronomers, regarded the defining problem of
sidereal astronomy as the spatial and kinematic arrangement of stars in the visible universe. The
Sidereal Problem  was then the search for order, trying to understand Kapteyn's two star streams
as the motions of particles in a spiraling vortex \citep{1993mwgs.book.....P}.

What of the spirals then?  Were they part of the stellar system, or external to it?  Russell
was aware of the problem, but to appreciate where they (or any unexpected observation) fit in
his larger scheme of things, we need to appreciate the state of the profession of astronomy at
the time.  This was a time of war, and for some, including Russell, there were questions of what
was critical for the future of the profession after the war.

The most contentious issue for the National Research Council (NRC)\footnote{Of the National Academy
of Sciences}, was what would be the nature of science after the Great War. In George Ellery
Hale's view, would international relations be re-established?  Who was in, and who was
out?\footnote{See \cite{1995.book.....K,kevles1971,kevles1968}.} Hale organized the deliberations around
disciplines, and asked Russell to take the lead for astronomers, concentrating on what the nature
of the discipline should be like.   As \cite{Russ-17} explained his position to Harvard's E.~C. Pickering in
November 1917,

\begin{quote}
[T]here are two sides of astronomical research, one of which has to do with
the collections of facts, and the other with their interpretation.
\end{quote}

The former was routine; the latter was not, and this was why, \cite{Rus-17} claimed, none of
Pickering's Harvard staff had been asked to write reviews for the NRC.  Pickering's own earlier
reports, he pointed out to his old patron, had already covered the routine aspects.  Now the
other side needed airing:

\begin{quote}
[I]t is upon studies of this sort that the future advances of any science must
very largely depend.
\end{quote}

Considering that it had been Pickering who made the vast holdings of Harvard spectra and
magnitudes available to Russell in past years, which led to his landmark study of the relations
between the characteristics of the stars, Russell's views could be thought of as insensitive.

\begin{figure}
\center{\includegraphics[scale=0.5,angle=-90]{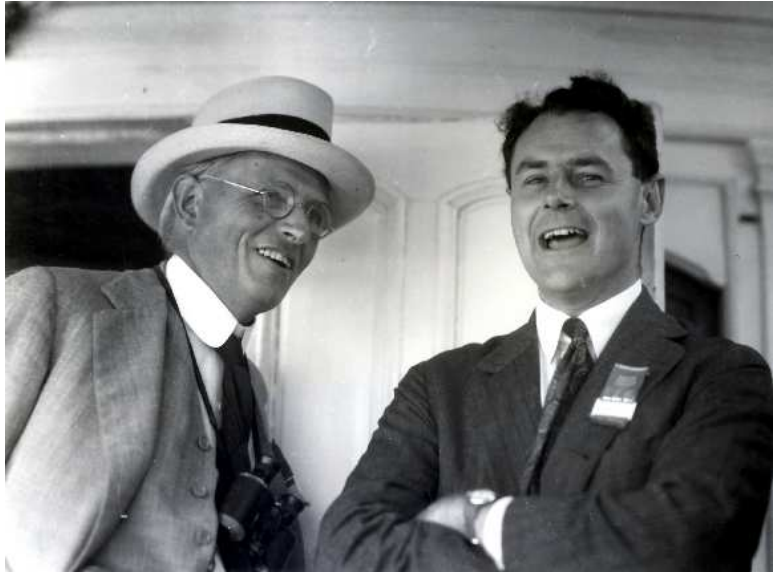}}
\caption{Starting in the late 1920s, Roger Lowell Putnam, the Observatory Trustee (r), was
receptive to Russell's (l) advice and intellectual patronage. Photograph by Clyde Fisher,
circa late 1920s, Neg. \#280362, courtesy Department Library Services, American Museum
of Natural History.}\label{devorkinfig01}
\end{figure}

Russell indeed gave highest marks to interpretation, but he was always very deferential
to those who collected the most interesting and useful observations, like Pickering, and indeed
like Slipher, for it was upon their work that his career was based.  His patronage of Slipher and
Lowell Observatory needs to be understood in this way (as does his patronage of other observatories like
Mount Wilson).  Russell's essay \citep{1920PA.....28..212R} had to wait till war's end, and there
he expressed his opinion very forcibly:

\begin{quote}
The main object of astronomy, as of all science, is not the collection of facts, but the
development, on the basis of collected facts, of satisfactory theories regarding the nature, mutual
relations, and probable history and evolution of the objects of study.
\end{quote}

When those facts were collected properly, and presented clinically, they had to be dealt with.
Indeed, for Russell, observation guided theory as much as theory informed plans for additional
observations.  This was a very new idea for American astronomers, one that few could assimilate.
In fact, ``Some Problems" was an appeal to observational astronomers to appreciate the power of
actually designing observational programs that answer specific questions.   Very much due to
Percival Lowell's problem-oriented style, the Lowell Observatory staff were more receptive than
most.

Russell's feeling was that one of the most important results of a wider sampling of spiral
radial velocities was to better determine solar motion:  best determined once radial velocities of many
nebulae are secured in all directions.  This was a classical application he shared with most others
who commented at the time.\footnote{See \cite{1982eua..book.....S}.}
So through 1916 to early 1917, at least, he
expressed little interest in the nature of that larger realm, though he was rather open to the
possibility that the nebulae were at great distances, even beyond the Milky Way. In fact, as early
as 1903, in a \emph{Scientific American} column, prior to the rise of the Chamberlin--Moulton encounter
theory and still feeling safe in a planetary system rendered common by the Nebular Hypothesis,
\cite{Rus-03} went so far as to speculate that, viewed from a distance of a million light years,
``our own
Galaxy would appear as a spiral or a ring nebula, something like the Great Nebula...But it must
be clearly borne in mind that the evidence available at present is too scanty to justify us in
making any definite statement..."
Thus for Russell in 1903, at least, the scale of the universe was not an issue.

He maintained this view, more or less.  In a series of lectures in the Princeton Chapel in
1916 and 1917, Russell pondered the ``Scientific Approach to Christianity" mainly to argue that
there need be no antagonism between ``Genesis and Geology" and that those at play were due to
mutual misunderstandings that he hoped his lectures would put to rest.  In one lecture, ``What
Nature itself looks like to the man of science," Russell speculated that some of the fainter spirals
could be upwards of 200,000 light years distant and that there was ``No evidence that we have
met with the limit." Echoing the emerging findings of his former graduate student, Harlow
Shapley, who was now at Mount Wilson, \cite{Rus-16A} wholly accepted that ``We are some 60,000 ly
from the center of the galactic system is now known\dots ."  He saw no inconsistency
between the growing evidence of the nebulae being beyond the Milky Way and of the vastness of
the Milky Way itself.\footnote{Shapley was coming to this distance estimate in late 1917. See
\citet[][p. 48]{1918PASP...30...42S}.}   Nor did he express any concern for divergent time scales in
Genesis and Geology at this time, evidently of a generation for which the ``gap theory" was no
longer necessary.\footnote{On the gap theory, see \cite{Moore-86}.}
But what of the motions of the spirals?

\section{Adjusting to Shifting Evidence}

Indeed, one can easily trace Russell's initial growing acceptance of the extragalactic
status of the spirals through his \emph{Scientific American} articles.   Despite his initial skepticism,
Russell was sufficiently impressed by June 1914 to encourage his \emph{Scientific American} editors to
invite Slipher to state his case for nebular motion and especially rotation, discussing its
importance for ``stellar and nebular evolution."  And by April 1915, \cite{Rus-15} felt that Slipher had
collected enough evidence for the phenomenon to devote most of his column to his work,
opening his discussion saying ``Among the most difficult of the yet unsolved problems of
astronomy is that of the distances, dimensions and nature of the nebulae."

Photography, Russell emphasized, had recently shown that the spectra of the spiral
forms of these objects resembled stellar spectra, wholly unlike the spectra of the brighter ``green"
nebulae.\footnote{At the time, astronomers still called diffuse nebulae, like Orion, green nebulae because
of the visual dominance of a greenish tinge caused by the oxygen emission spectrum.  White
nebulae, like spirals, displayed continuous spectra.  An explanation is given in:
\cite{2011arXiv1108.4864T}. See also \citet[][p. 91]{1932PASP...44...89D}.} This meant they were either
single stars cloaked in gas and dust, or assemblages of stars at vast distances. Russell then
introduced Slipher's accumulating observations of spiral radial velocities,
now confirmed by ``others" that showed them to be in the range
of many hundreds of miles per second.

What was important for Russell was that the average of the radial velocities for the
spirals was some 25 times the average for stars.  If there was such great motion in the line of
sight, one would expect there to be some observable translational motion as well. But \cite{Rus-15}
then noted that none of these nebulae showed any sensible proper motions, none exceeding 0.1
arc second per year, ``though one out of every 4 stars exceeds this limit..."

Taking the planetary nebulae observed by the Lick astronomers, and the spirals observed
by Slipher, as two classes, their measured radial velocities indicated statistically even higher
space velocities, which Russell translated into AUs/year. The planetaries travelled 15 AUs/year
and the ``white spirals"  were travelling an astounding 120 AUs per year or more.   Given this
estimate, and the vanishingly small proper motions of these diffuse objects, or their stellar cores,
they had to be at least 8 to 10 thousand light years away, but more likely in the ten to twenty
thousands of light years.
``The dimensions which the nebulae must have, to appear as big as they do at
such distances, are astounding."

Comparing Ejnar Hertzsprung's recent estimate of the distance to the Small Magellanic Cloud,
some 30,000 light years, \cite{Rus-15} further speculated that the distances to those unresolved
nebulae must be even greater:

\begin{quote}
...it may well be that we are on the brink of an expansion of our conception of the extent of the
universe almost comparable to that which resulted from the first measurements of the distances
of the nearer stars."\footnote{\citet[][p. 204]{1913AN....196..201H} converts a parallax of
0.0001 seconds of arc into a distance of 3000 light years, where it would be closer to 32,600 light
years. \cite{Rus-15} related it as 30,000 light years.}
\end{quote}

In September \cite{Rus-16} hailed V.~M. Slipher's observations of the great velocities of
the spiral nebulae, which, along with the work of the other ``Pacific Coast Observatories...has
so greatly widened, at a single bound, the limits of distance at which our investigations may
operate."

But by late 1917, influenced somewhat by Shapley, who was excitedly reporting Adriaan van
Maanen's provocative conclusions about the rotation of spirals, \cite{Rus-17} started reversing his
views.  Responding to Shapley in November, Russell accepted van Maanen's observations, which
immediately threw doubt on the Island Universe theory:  ``But if they are not star clouds," Russell
mused, ``what the Dickens are they?" \footnote{Also quoted by \citet[][pp. 36-7]{1982eua..book.....S}.}

        Russell never overtly criticized van Maanen's results. Indeed, since 1916 he had gained
high respect for his astrometric abilities, convinced that van Maanen's deft re-analysis and
refinement of stellar parallaxes, especially his analyses of internal errors, was a very strong step
forward in that delicate art.  He looked upon van Maanen as a new breed of astrometrist, more
adept at mathematical analysis than his predecessors, like Benjamin Boss or even J.~C. Kapteyn.

        As much as Russell evidently respected van Maanen's abilities, his early results for
rotation in spirals were so extreme that Russell remained cautious, despite Shapley's promotional
efforts.  However, in October 1920, after van Maanen reported directly to Russell that he had
found much the same rotational properties for another spiral, M 33, Russell expressed both
excitement and great relief:

\begin{quote}
I have been waiting with great interest to see what you would get on additional
spiral nebulae and this confirmation of your early discovery is most gratifying.
\end{quote}

\cite{Rus-20} looked forward to more confirmations from Mount Wilson; in conjunction with
Pease's spectroscopic efforts, he encouraged van Maanen to obtain distances to
these objects, so that they all could proclaim:

\begin{quote}
Good-bye to Island Universidus!
\end{quote}

Privately to van Maanen, he was even willing to play down the nagging question of the
intrinsic brightnesses of various novae events seen in the spirals by Curtis at Lick,
admitting that they were dim compared to galactic novae.  They might well be intrinsically
dim, Russell speculated, which he felt was well within bounds for discounting this odd class of object.   

Thus in the two years between the time Russell first drafted the essay for the NRC,
and when it was finally published in October 1919, and reprinted in \emph{Popular Astronomy} in 1920,
van Maanen's rotational proper motions had already raised serious questions about the existence
of a hierarchy beyond the Milky Way, questions Russell was evidently happy to raise.  Russell
clearly wanted van Maanen's results to be confirmed, and so in his essay he went so far as to
suggest, without mentioning de Sitter, that the distances the light had to travel from the
spirals might also be causing the effects Slipher was seeing.   

Shapley, of course, was wholly convinced by his own work on the distances to globulars
and their association with the Milky Way, but it was not the deciding factor for Russell.  Even
without the confirmation that would come in October 1920 from van Maanen, Russell exhibited
his enthusiasm in his National Academy essay, concluding for the moment that the distances to
the spirals, and their intrinsic sizes, could not be ``extragalactic" largely owing to van Maanen's
remarkable proper motion measurements.

Russell's rejection of the extragalactic framework came after, as Robert Smith and other
historians have related so well, Slipher's continuing efforts, now bolstered by Pease's spectra with
the 60 inch, had won many astronomers over.  There was strong evidence that spirals were not
solar systems in formation: they were not stars. To Hertzsprung and others these speeds meant
that the spirals could not be bound to the Milky Way.  International acclaim soon followed: In his
Presidential address at the Annual General Meeting of the Societ\'{e} Astronomique de France in
September 1917, even in the depths of war, the Count \cite{1917Obs....40..327D} rejoiced that
much work had been done recently, especially noting Slipher's spiral radial velocities which ``give
a further proof of the independence of the spiral nebulae of our Universe."

In the early 1920s, therefore, Russell remained an agnostic over the meaning of Slipher's
velocities.  He never doubted the observations themselves any more than he doubted van
Maanen's measurements; at every turn, Russell indicated that Slipher's careful work had to be
factored into any discussion, for just as van Maanen had proven to be a pioneer in refining
astrometric technique, Slipher had certainly done so for spectroscopy.  In 1920 Russell became
especially excited by Slipher's recent work on the Orion stars \citep{1919PASP...31..212S},
writing to Adams that it ``looks to
me as if the luminescence of the nebula was due to some influence emanating from the hot stars
within it..."  This was another piece of evidence that ``the nebular lines are produced by known
atoms, under unknown conditions of excitation" \citep{Rus-20b}.
Russell never doubted Slipher's observations, and supported them through speculating about what they
revealed about the stars, something Slipher rarely if ever tried to do.

        Several months before Hubble's revelatory announcement of the distance to M31 based
upon Cepheids of some 285,000 parsecs, Russell gave a series of popular
lectures in February 1924 at the University of
Toronto.  During the 14th and final lecture, on the ``other classes of great nebulae" - those of
spiral and globular forms, he very much followed the evidence van Maanen offered for 7 spirals
showing they were in rapid rotation, throwing off material, very much in line, Russell believed,
with James Jeans' theoretical models. Given all the evidence now at hand, Russell concluded, the
spirals themselves could not be more than some 10,000 light years distant, which was not large
compared to the size of Shapley's galaxy: ``... if [van Maanen's] measurements are right, these
things cannot be anything at all like as big as the Milky Way."  The Andromeda nebula, however,
was for \citet[][p. 259]{Rus-1924} still one of the most
``remarkable objects known to astronomical science." 
Given its enormous size and luminosity, and its apparent violent motions throwing off matter, it was a
clue to formative processes in his universe.

        Russell's highly convoluted presentation appears in hindsight to be due to his attempt to
reconcile the wild observations at hand.  But in fact he was most influenced by how, in his mind,
Jeans' evolutionary model provides ``a beautiful picture of the spiral nebulae" as evidence once
again of his cherished nebular hypothesis.  ``It looks as if we have stages of evolution shown"
Russell believed; Jeans could even account for the lumpiness of spiral arms.  But Russell
admitted that a very recent paper by Jeans which ``came out only a week or so ago shows that
when we want to be quantitative it is much more difficult." Jeans' latest model suggested, in fact,
that spirals become more tightly wound
with time.\footnote{\cite{Jeans1923MNRAS..84...60J} discusses the tightening of the arms.}
\citet[][p. 265]{Rus-1924} characteristically reminded his audience that
science was tasked with how spirals ``got into their present shape" but was not
equipped to say: ``...how they got there, because of that we do not know
anything at all. It is no business of
science to explain how matter got into space; only supposing there was matter in space, what
would be the history of it." As \cite{Rus-1924} concluded, ``Only a portion of
Genesis is accessible to the realm of science."

Everything changed, of course, by the end of the year and Hubble's work announcing the
distance to M31.  Russell heard the news first from James Jeans when they happened to meet in
New Haven sometime in the fall, and then later in November. Finally, Hubble reported directly in
December and \cite{Rus-24} responded heartily: ``They are certainly quite convincing...It is a beautiful
piece of work..."; Russell urged Hubble to send his paper to the Washington meetings of the
American Association for the Advancement of Science in February 1925, if for no
other reason than his paper would no doubt ``bang that
\$1000 prize." Russell read Hubble's paper at that
meeting\footnote{\citet[][pp. 111-120]{1982eua..book.....S} and
\citet[][Chapter 5]{1976mdg..book.....B} have explored in some detail the impact
of Hubble's first paper on Cepheids on the acceptance of van Maanen's rotational
measurement work.}
saying nothing of the implications this new result had on van Maanen's measurements.

Jeans and Russell did meet in New Haven in August 1924, presumably because there
were two major meetings starting with the American Astronomical Society, held at Dartmouth
and timed specifically to coincide with the special annual meeting of the British Association for
the Advancement of Science in Toronto.  Russell reported on both in his \emph{Scientific American}
column, highlighting the recent studies of Yale's E.~W. Brown on the osculating orbits of ``particles
composing the arms" in spiral nebulae.   Both Jeans and Russell later acknowledged that they
spoke of Hubble's observations and their implications for Jeans' theory of spirals in New Haven.
And after Jeans returned to England in late October he wrote to Hubble based upon his
conversations with Russell, sending a copy to Russell, setting out points that Jeans believed
supported Hubble's new estimate of the distance to M31.  Jeans made his points rather
dramatically.  Using Hubble's new distance to M31, and F. Pease's spectroscopic observations of
its rotational motion, ``we found a period of rotation of 17,000,000 years."  Then \cite{Jea-24}
speculated to Hubble:
``Suppose that Van Maanen had measured motions in M.31 and had announced
this period.  If I had calculated the parallax by the method I used for M.101 on
p. 217 of my ``Cosmogony" (assuming, as I did there, that the condensations are
3$\arcsec$ of arc apart) the result I should have obtained would be
(17,000,000/85,000) $\times$  5/3 $\times$ 1,000 parsecs = 333,000 parsecs.
Which is near enough (rather too near!) to your estimate."

At first glance, it seems as if Jeans employed Hubble's distance post hoc  but in fact, he found it
by a very independent method: using Eddington's just announced mass/luminosity law for stars.
Using the observed brightnesses of the average condensations in the nebula, their angular
distances apart, and Pease's radial velocities, he found the distance to M31 to be 337,000
parsecs, employing Hubble's estimate of the absolute magnitude for the Cepheids of M = -4.  To
which \cite{Jea-24} again noted: ``which again is near (too near) to your estimate"
\footnote{Also noted in \citep[][p. 120]{1982eua..book.....S}.} and bluntly concluded:
``In view of these calculations, I feel no further suspicions of the period-luminosity
law, or of my physical interpretation of the condensations -- on the other hand, I
fear van Maanen's measures have to go."

Jeans was now able to show that Eddington's mass-luminosity law, as well as the period-
luminosity relation, combined with Hubble's distances, supported his own concept of the nature
of the spirals.\footnote{Jeans described his thinking in greater detail in \cite{1925MNRAS..85..531J}.} 
Here \cite{1925MNRAS..85..531J} increased his earlier estimate of the distances to spirals like M31
(originally 5000 light
years), in light of Hubble's and Shapley's work, to some 950,000 light years, and concluded: ``If
the same correlation [Eddington's M/L relation] is assumed to hold for the stellar condensations
in a nebula, we can dispense with v. Maanen's measurements altogether and (in theory at least)
determine the distance of the nebulae from the observed stellar magnitudes of its
condensations." \cite{Rus-24b} agreed, writing Jeans hastily in early
November 1924 that ``the agreement of the new data with your theory is beautiful..."  Russell did
not elaborate, however, saying there was nothing he could add off the top of his head.

\begin{figure}[ht]
\center{\includegraphics[scale=1]{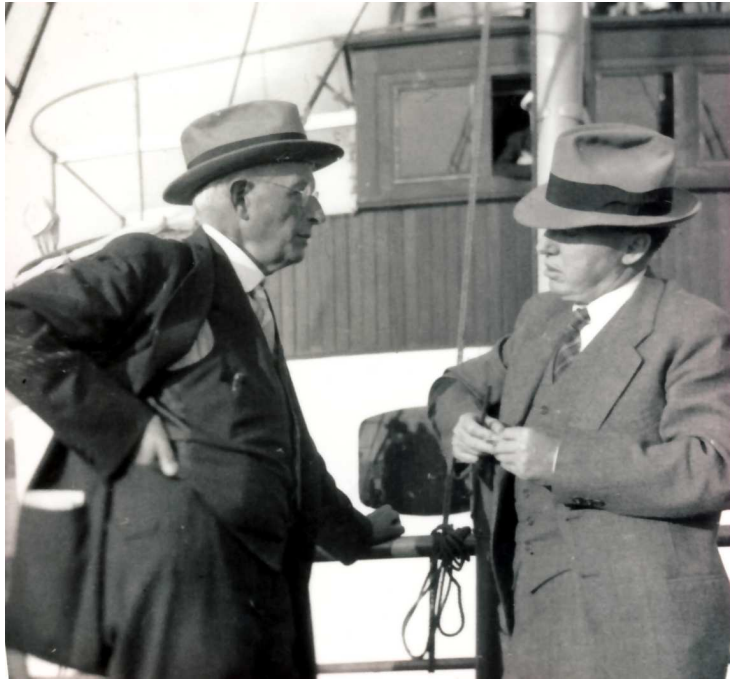}}
\caption{Russell and Shapley remained life long colleagues and friends.
Here they are captured aboard a steamer tour at the 1938 International
Astronomical Union General Assembly in Stockholm. Photography by
Dorothy Davis Locanthi, Courtesy E. Segr\`{e} Visual Archives,
American Institute of Physics.}\label{devorkinfig02}
\end{figure}

Throughout 1925, Jeans explored the ramifications of Hubble's distances, clearly
delighted with the new framework.  Russell attended to his solar and stellar spectroscopic
studies, experimenting with the emerging explanatory framework from quantum physics to find
useful links between the laboratory and the stars.  He could not have relished Jeans' enthusiasm
for Eddington's mass-luminosity relation, of course, because it meant that, in Russell's own
words \citep{Rus-26}, it put his theory of stellar evolution ``in a state of chaos."
Later \cite{Rus-28a} 
admitted openly that ``[t]he path of this idea is strewn with the wreckage of abandoned theories,"
and that the field was at an impasse. In like manner, even though Russell could not
have been comfortable with the implications of Hubble's vast distances, when he reported on
them for \emph{Scientific American} in March 1925, he also credited Shapley's distances to globulars and
hence his estimate of the size of the Milky Way as having ``expanded our existing ideas of the
universe ten-fold." He did this, evidently, to assert that  Shapley's universe
was still viable (see Figure 2).  He now fully accepted the vast distance
to M31, admitting that it ``leads to conclusions which are
enough to make even a case-hardened astronomer gasp."  M31 was now, at Hubble's distance of
a million light years, at least 35,000 light years in diameter, which would have made it larger
than the Milky Way on Kapteyn's old model. But, \cite{Rus-25} hastened to point out, it was still only
1/10th the size of the Milky Way given Shapley's work.  This lingering attachment to
the idea of the Milky Way as being somehow special provides insight into the nature of Russell's
personal universe.

\section{Russell's Universe}

        In addition to his scientific writings, as noted above, Russell actively wrote and lectured
on science and religion for a widening audience, starting on campus after he attained the status
of a Princeton professorship in 1908, promoted to full professor in 1911 and Director of the
Halsted Observatory in June 1912.   He espoused liberal theological views, reflecting his
upbringing as the eldest son of a New School Presbyterian pastor, and by the 1920s took on the
role of a modernist, intent on arguing that there was much theology could learn from science in,
what he was coming to accept evermore, was an age of relativism.  Russell's efforts to explore
the interface between science and religion thus promises further insight into his reaction to the
concept of an expanding universe.

In 1925 he was invited to Yale to deliver the annual set of Terry Lectures, which he titled  
``Fate and Freedom."  Keenly aware that Presbyterian theological orthodoxy was still struggling to
kill the liberalism of New School reformers, Russell asserted his modernism, proclaiming that
``The old hope of finding final and perfect statements of the truth about a simple universe has
fled." In its place was a striving for closer and closer approximations to the truth, ``a steadily
increasing accuracy of approximation in the description and interpretation of an incredibly and
magnificently complex universe." In its constant striving, science had freed itself from absolutes,
whereas theology had not \citep{Rus-27,Lon-91}.

During these years, as he gained prominence, he maintained a lengthy correspondence 
with inquirers seeking illumination about God from the world of science.  In late November 1928,
he responded to the inquiries of a reader of ``Fate and Freedom," the Beacon Hill doyenne 
Margaret Deland, an American novelist and poet and particularly articulate inquirer. Deland,
inspired by her reading, hoped her views as a ``philosophically uneducated" admirer, in the spirit
of Moliere's ``average mind," might be of some use to the Princeton savant.  The revelation she
wanted to share, without attribution, was ``that in this universe of evident order and purpose,
God is still `becoming'."   Only in this way, she felt, could she understand that ``our imperfections
may be steps in His (I wish I didn't have to use the personal pronoun!) process of
`becoming'"\citep{Del-28}. Although she makes no mention of Alfred North Whitehead's influence 
here, especially his widely read 1925 ``Science and the Modern World" there are similarities with
what he later espoused as ``process philosophy."  If there is a connection and insights to be had,
I leave them for others.  Thanks though are due to Matt Stanley for raising this question during
the conference.

\cite{Rus-28b} responded at length, clarifying his position, and his universe:

\begin{quote}
I suppose it is because I am a physicist that I find it hard to think of God as
``becoming." Of course, we can none of us hope for a full understanding of the
Power behind the universe. But those of us who study the universe itself, at least
from the material side, do not find evidence of progress at the heart of things.
There is plenty of development and evolution of individual parts of the universe,
from stars to souls; but these are all particular cases of the operation of
invariable laws of nature.
\end{quote}

Russell took the laws of physics as invariant in a universe that, on the whole, was unchanging,
though the things in it do change. \cite{Rus-28b} thought of his ``God as manifesting Himself rather than
evolving."

Just over a month later, in early January 1929, Russell's unchanging universe was again 
challenged.  Soon after he reached Pasadena in what had become a yearly trek cross country to
advise and counsel, he heard the local news about what Hubble was now up to.  In a long letter
to Lowell astronomer Carl Lampland, who was then in residence at Princeton in a scheme Russell
created that hopefully would better acquaint Lampland with mainstream astrophysics,
\cite{Rus-29} reported that Hubble:
\begin{quote}
... has definite evidence, at last, that the radial velocities of non-galactic  
nebulae increase with distance ... the explanation ``may" involve a ``de Sitter
Universe" or some new mathematical invention - it is too early to say...
\end{quote}

Russell's expression ``at last" highlights the fact, as Robert Smith has well shown
\citep{1982eua..book.....S}, that the velocity/distance relation was very much in the air in the 1920s. 
But he mentioned it only in passing, devoting most of his long letter reporting on stellar
astronomy and new instrumentation at Mount Wilson.

Within a few months, as Russell gathered his thoughts for his \emph{Scientific American} column,
the idea of an expanding universe did not sit too well, although, as usual, the observed facts
were not in question.   It was the implication.  In a column titled  ``The Highest Known Velocity"
Russell made much of Milton Humason's tedious work, taking exposures over several nights, of
between 33 and 45 hours duration, and the resulting ``very pronounced distance-effect."
Regarding it all as very ``strange" Russell asked ``what does all this mean?"  Could this be a
``peculiar form of the theory of relativity suggested by .... de Sitter" or was it something wholly 
different.  Conversations with Richard Tolman helped Russell appreciate how a de Sitter universe
could look like a ``now familiar" Einstein universe on small scales, but the two differed, as he
reported to his readers, ``in a larger view, [where] space and time have unexpected properties."
On small scale, two particles at rest would remain at rest save for a mutual Newtonian attraction.
But at vast distances, such as those between the nebulae, they would recede from one another
and scatter ``ever more and more widely and recede from one another faster and faster."  Even
though de Sitter's model literally predicted what Hubble had found, \cite{Rus-29B} concluded that it was
``premature" to adopt de Sitter's ideas: ``The notion that all nebulae were originally close together
is philosophically rather unsatisfactory..."

Russell's reaction was distinct only insofar as he did not question the data.   Others
certainly did.  At about the same time that Russell wrote his column, V.~M. Slipher, along with his
friends at Mount Wilson, notably John C. Duncan and Walter Sydney Adams, questioned the
reality of Hubble's and Humason's observations.  The fact that two nebulae in the Coma group
had velocities differing by some 3000 km/sec made \cite{Dun-29} skeptical; he happily reported to
Slipher that Adams felt this discrepancy was throwing ``doubt on the de Sitter theory and making
the whole situation more interesting."

Hubble's continued confirmation and strengthening of the velocity-distance relation had
to be faced by astronomers, one way or another.  I'll end by discussing briefly how Russell dealt
with it through the rest of his career.

The first edition of Russell, Dugan and Stewart's 2-volume Astronomy textbook was
completed in 1926 and published in 1926 and 1927.  A second edition appeared in 1938
\citep{1938Rus..book.....R}. In the
first edition he outlined Slipher's observations in some detail, and, not questioning the
observations themselves, still asked what it all meant: ``Whether this represents a real scattering
of the nebulae away from this region where the sun happens to be is very doubtful," Russell
asserted in a smaller pitched paragraph (a tactic he used to highlight speculative sections),
adding that some form of ``the generalized theory of relativity" that space is bounded and
``returns into itself"  might explain the spectral shifts.\footnote{\citet[][p. 850]{Rus-27}}
In the second                            
edition a decade later, Russell voiced this same concern in his chapter on ``Nebulae," but in a
later chapter on the ``Evolution of the Stars" that had been extensively rewritten, he included a
long passage on the problem of reconciling various time scales: the earth, the stars, and the
universe.  Here Russell did mention Hubble's 1929 work in a supplementary section, but it was
highly terse, clinical and with little elaboration.  He made no mention of Lema\^{i}tre.

Russell also wrote popular essays on Einstein's theories of relativity and toyed with the
visible consequences of intense gravitational fields for his \emph{Scientific American} audience.  Even
though Russell never tried to contribute to general relativity or theories of the expanding
universe, he was fully aware of the potential impact both might have on modern
astrophysics, writing about them
with clarity and style, making them accessible to the general reader and
rank-and-file astronomer
alike \citep[][p. 321]{Kaiser1998SHPS...29...321K}.\footnote{As Kaiser and
others discuss, in the 1930s,
Princeton harbored more interest in General Relativity than most American
campuses, yet it was not a central topic in physics until the 1960s.}

Russell's most revealing response to the expanding universe was how he used it to
restore the possibility that we are not alone in the universe.  Even though he initially was
repulsed by the idea of the nebulae being closer together in the distant past, a rather surprisingly
rash statement to make in a \emph{Scientific American} column, once he had the benefit of years and
time for rationalization, he realized that in fact he could see something useful in the framework.

Deprived of the Nebular Hypothesis by the popularity of the Chamberlin--Moulton
encounter theory, Russell always looked for ways to reinstate
the theory (see \cite{1978JHA.....9...77B,1978JHA.....9....1B}).
 On its influence on the ubiquity of planetary systems, see
\cite{2001low..book.....D}. The expanding universe gave him an opportunity.  The solar system
was formed, he argued in his 1935 University of Virginia lectures ``The Solar System and its
Origin," at a time when the universe would have been much smaller than it is now, at a time
when collisions were more probable.  He thought more \citep{1935ssio.book.....R},
in fact, of the idea that the solar system was formed at a time when, some two billion years ago:
\begin{quote}
...a very great event occurred. This would have been the time {\it par excellence} for
encounters of all sorts.  Or it may be that a cosmic New Deal occurred, and that, just afterwards,
matter was distributed more widely, but more thinly, through space, to settle down
into the stars.\footnote{pp. 137-8}
\end{quote}

This was the approximate epoch when the presently observed expansion of the universe had its
start.  It also agreed, he then thought, with the radioactive age dating of rocks, the tidal
evolution of the earth-moon system, and was well within the possible lifetimes of the stars. In
this distant epoch, the probability of encounters may have been far greater than today.  Not only
were stars like the sun larger than at present in his lingering view of the course of stellar
evolution, but the universe was far more compact.  Russell speculated that at this distant epoch,
encounters and collisions could have been a common occurrence.  Following a suggestion he
attributed to Willem de Sitter, he was willing to speculate \citep[][p. 25]{1935ssio.book.....R}
that the stars themselves pre-existed this moment of crisis in the universe, many coming through it
unscathed. Collisions would have been more frequent in the past,
relieving all of the encounter theories of the burden of making mankind feel alone in the
universe.   Envisioning a time when ``all the matter of the Universe was tighly packed together,
perhaps in the one great Atom, which forms the starting point of Lema\^{i}tre's hypothesis,--which
deserves no less respect because of its picturesqueness," \citet[][p. 138]{1935ssio.book.....R}
thus ended his lectures on a hopeful note for the ubiquity of life in the universe. In commentary in
\emph{Scientific American}, popular speeches, and radio talks, Russell harbored a deep desire to believe
that human beings were not an accident in the universe. For a talk at Bowdoin College in Maine
in April 1941, Russell relied on the humor of Gilbert Chesterton, who once defended his belief
that he had been born, even though he had no recollection of it and could not test it.

Russell always remained optimistic that somehow, in time, with effort, discrepancies
would be reduced, and consensus would come, and though consensus would come a deeper
appreciation of the universe we live in.  He delighted in describing the process of science as
building on what was, for him, ``a tissue of approximations," as he described it in his 1937 AAS
Presidential address.\footnote{Quoted in \citet[][p. 246]{2000hnr..book.....D}.}
There was great heuristic value in
the process, as long as physical insight was not lost, and the clues that observational data
provided were faithfully followed up and tested for veracity, but never swept under the rug.
Thus, throughout this period when others looked askance at the disparity in the time scale for
the age of the universe computed from Hubble's recessional velocities, compared to the ages of
all the things in it as derived by radioactive determinations of the age of the earth, the ages of
stars allowed by nuclear energy sources recently uncovered by Bethe and Marshak, the time
scales from dynamical studies of binary stars and star clusters, Russell searched as usual for a
silver lining.  In his May 1940 James Arthur Foundation lecture at New York University, Russell
could at least hope that the inference of Hubble's evidence,  best described, he thought, ``a little
speculatively" by the Abbe Lema\^{i}tre, could make it possible to have a universe filled with
planetary systems:

\begin{quote}
The principal attraction of this scheme is that it pictures a short but tumultuous time in the early
days of our present universe during which all sorts of things which can never happen now might
have occurred, such as the origin of the planetary system and ... the formations of the heavy
radioactive atoms.\footnote{\citet[][p. 16]{Rus-49}. Russell did wonder privately if the heavy
elements were older than the stars, in letters to van Maanen and Shapley in 1939.
See \citet[][p. 254, ref. 83]{2000hnr..book.....D}.
For a contemporary review of time scales, see \cite{1946MNRAS.106...61B}. Good
secondary sources that address one or more aspects of the time scale problem include
\cite{1978JHA.....9...77B,2001low..book.....D,1991esss.book.....H,1975.book.....B}.}
\end{quote}

Russell acknowledged that if this were so, then the ages of everything
had to be limited to less than 2 billion years time.  Even so, in concluding his lecture,
Russell rejoiced that science had at
least succeeded in showing that the time scale of the universe can be measured in billions of
years, not millions, ``and in very few of them" as concluded from ``four independent lines of
evidence." \footnote{\citet[][p. 29]{Rus-49}. Russell identified the ``outstanding difficulties"
remaining as the persistence of giant stars and the premature appearance of white dwarfs.
This was at a time before Baade's populations were articulated, and when giants were
still considered, by Russell at least, to be young stars, in spite of the initial evidence
provided by Str\"{o}mgren, Kuiper and Gamow. See \cite{2006JHA....37..429D}.}

But this was as far as he was willing to go.  He ended his essay acknowledging that there
was a ``widespread desire" that this expanding universe might someday reverse itself and
contract, leading to ``some cyclical restoration of activity."  But he was not in sympathy with this
view, expressing surprise that it had such a ``strong aesthetic hold" on so many people. Here
Russell enjoyed a bit of Eddingtonian wordplay to make his point:
``I am an evolutionist, not a multiplicationist. It seems so tiresome to me doing
the same things over and over again." \footnote{\citet[][p. 30]{Rus-49}}

Indeed, this is the same Russell who in 1914 chided a critic's rejection of his penchant for making
inferences based upon statistical behavior and not direct measurement: ``A hundred years hence
all this work of mine will be utterly superseded: but I am getting the fun of it now"
\citep{Rus-14}.\footnote{Doolittle had rejected Russell's technique of hypothetical parallaxes.}
So even while he could marvel at Slipher's high radial velocities,  van Maanen's proper motions or
Hubble's vast distances, following the observations to see where they lead, he knew only too
well, and was comfortable with the fact, that the implications one might draw from them were fleeting.

Nevertheless, in his last known commentary on the implications of the expanding
universe, in a foreword to the English edition of Lema\^{i}tre's ``The primeval atom,
an essay on cosmogony," Russell kept his distance from the question. Describing
it as an exercise in ``pure intellect;" as an hypothesis and not a theory, he felt
it ``goes far beyond the limits of present proof."  Indeed, not apparently aware
of, nor concerned with, the recent prediction of a redshifted remnant by Ralph
Alpher and Robert Herman in 1948, nor making any effort in his foreword to
explicitly link the theory to the observed redshifts themselves and that the universe must be
expanding, Russell slyly regarded Lema\^{i}tre's retrospective essay as  ``...a
fascinating view of the birth and growth of a noteworthy hypothesis regarding
the origin of the material universe." The only reference Russell made explicitly
to observational evidence dealt with the recent detection of high-energy cosmic
rays which emulated Lema\^{i}tre's ``atom." So as he did throughout much of his career, Russell held highest evidence
from physical observation, whether or not it tested his preconceptions. Just why
he was asked to provide the foreword, and not someone closer to the subject
matter, indicates his prominence in the field of professional astronomy in that
day.\footnote{\citet[][pp. v-vii]{Rus-50}}

\acknowledgements 
This research has made use of NASA's Astrophysics Data System Bibliographic
Services, and the kindnesses of the U S Naval Observatory Library. The author
has also benefited from comments by Robert Smith, Michael Neufeld and
Joseph Tenn, as well as by the editors.

\bibliography{devorkin}

\end{document}